\documentclass[fleqn,usenatbib, letters]{mnras}

\usepackage{newtxtext,newtxmath}

\usepackage[T1]{fontenc}
\usepackage{ae,aecompl}

\usepackage{graphicx}	
\usepackage{amsmath}	
\usepackage{amssymb}	

\title[LB-1 reconsidered]{Not so fast: LB-1 is unlikely to contain a 70 $M_{\odot}$ black hole}

\author[K. El-Badry]{
Kareem El-Badry$^{1}$\thanks{E-mail: kelbadry@berkeley.edu}
and Eliot Quataert$^{1}$\\
$^{1}$Department of Astronomy and Theoretical Astrophysics Center, University of California Berkeley, Berkeley, CA 94720 \vspace{-0.2cm}}

\date{Accepted to MNRAS \vspace{-0.5cm}}

\pubyear{2019}

\begin{document}
\label{firstpage}
\pagerange{\pageref{firstpage}--\pageref{lastpage}}
\maketitle

\begin{abstract}
The recently discovered binary LB-1 has been reported to contain a $\sim$\,$70\,M_{\odot}$ black hole (BH). The evidence for the unprecedentedly high mass of the unseen companion comes from reported radial velocity (RV) variability of the H$\alpha$ emission line, which has been proposed to originate from an accretion disk around a BH. We show that there is in fact no evidence for RV variability of the H$\alpha$ emission line, and that its apparent shifts instead originate from shifts in the luminous star's H$\alpha$ absorption line. If not accounted for, such shifts will cause a stationary emission line to appear to shift in anti-phase with the luminous star. We show that once the template spectrum of a B star is subtracted from the observed Keck/HIRES spectra of LB-1, evidence for RV variability vanishes. Indeed, the data rule out periodic variability of the line with velocity semi-amplitude $K_{\rm H\alpha} > 1.3\,\rm km\,s^{-1}$. This strongly suggests that the observed H$\alpha$ emission does not originate primarily from an accretion disk around a BH, and thus that the mass ratio cannot be constrained from the relative velocity amplitudes of the emission and absorption lines. The nature of the unseen companion remains uncertain, but a ``normal'' stellar-mass BH with mass $5\lesssim M/M_{\odot}\lesssim 20 $ seems most plausible. The H$\alpha$ emission likely originates primarily from circumbinary material, not from either component of the binary.
\end{abstract}




\section{Introduction}
Characterization of the Milky Way's stellar-mass black hole (BH) population is a primary aim of numerous stellar surveys. As part of a search for stellar-mass BHs with normal-star companions, \citet{Liu_2019} recently identified an intriguing BH-candidate (``LB-1'') consisting of a B star with an unseen companion. The B star is on a nearly circular orbit with $P=78.9\pm 0.3\,\rm days$ and velocity semi-amplitude $K_B=52.8\pm 0.7\,\rm km\,s^{-1}$, yielding a mass function (which represents the absolute minimum companion mass compatible with the orbit of the star) of $f(m) = 1.2\,M_{\odot}$. If the B star's mass is taken to be $M_B \approx 8.2\,M_{\odot}$, as \citet{Liu_2019} estimated from a model fit to the stellar spectrum, this corresponds to a lower limit on the companion mass of $M_2\gtrsim 6.3 M_{\odot}$.

However, \citet{Liu_2019} argue for a companion mass much higher than this, $M_{2}=68^{+11}_{-13} M_{\odot}$, which implies a low inclination ($i\approx 16$\,deg). The spectrum of LB-1 contains both the expected absorption lines of a B star, and a strong H$\alpha$ emission line, which the authors interpret as originating from an accretion disk around a BH. They find the H$\alpha$ line to be RV variable, with velocity semi-amplitude $K_{\rm H\alpha}=6.4\pm 0.8\,\rm km\,s^{-1}$, and interpret the ratio of the velocity amplitudes of the absorption and emission lines, $K_B/K_{\rm H\alpha}\approx 8.2$, as representing the mass ratio, $M_2/M_B$. Taking $M_B \approx 8.2 M_{\odot}$, this yields $M_2 \approx 68\,M_{\odot}$.

The proposed BH mass is more than double that of the most massive BHs in dynamically confirmed X-ray binaries \citep[e.g.][]{McClintock_2006} and is also more massive than the most massive pre-merger BHs thus far discovered via gravitational waves \citep{Abbott_2019}. The existence of a 68\,$M_{\odot}$ BH formed by stellar collapse is unexpected on theoretical grounds, because would-be progenitors of BHs with masses between about 50 and $130\,M_{\odot}$ are predicted to lose mass and then explode via pulsational pair-instability supernovae \citep[e.g.][]{Woosley_2017}, leaving behind either lower-mass BHs or no remnants at all. The novel and unexpected nature of the object has already spurred several theoretical investigations into possible formation channels for such a binary \citep[e.g.][]{Belczynski_2019, Shen_2019, Groh_2019}.

\begin{figure*}
    \includegraphics[width=\textwidth]{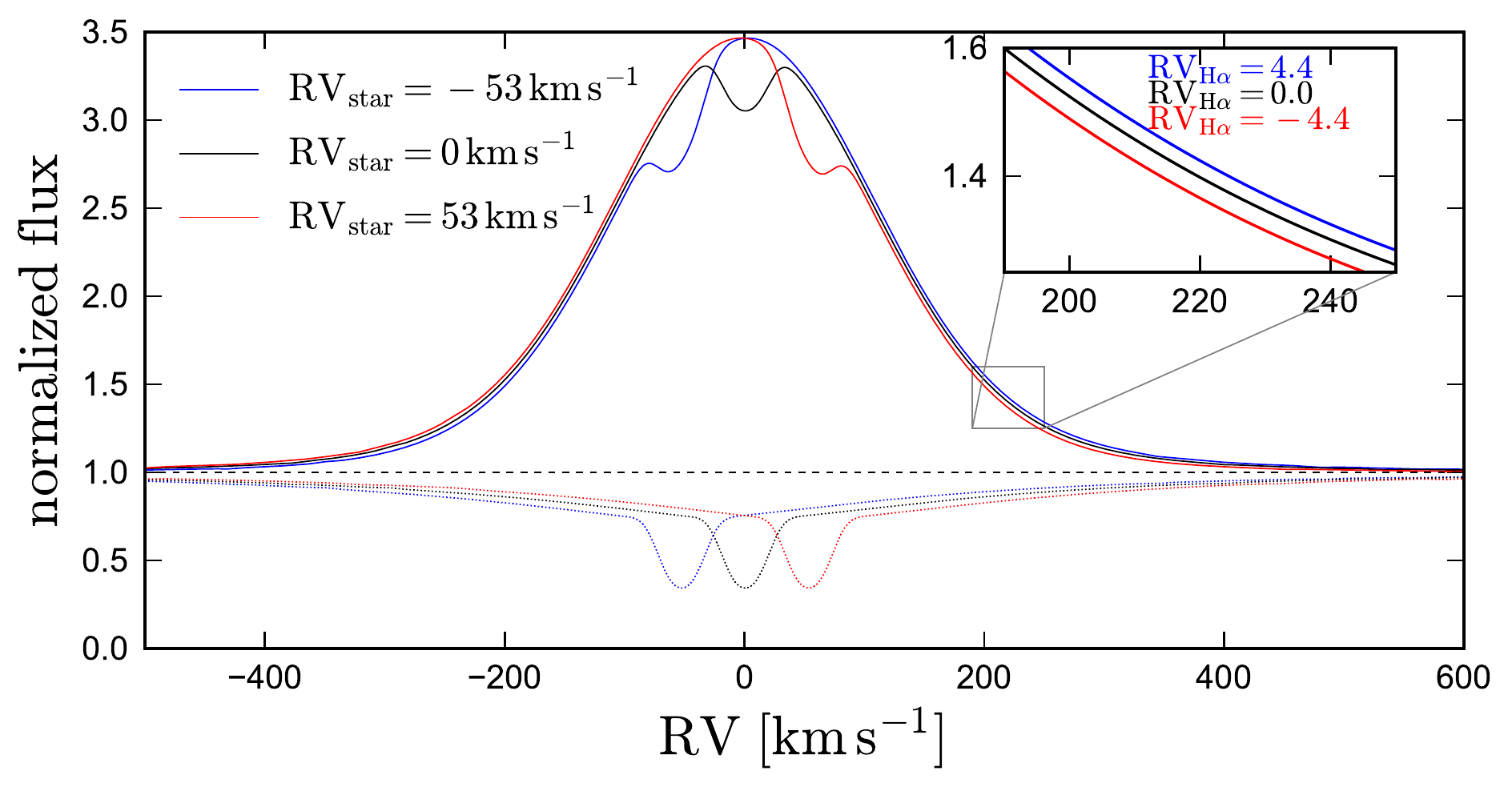}
    \caption{Schematic illustration of the effects of an RV-variable absorption line on the apparent velocity of a stationary emission line. The velocity of the emission line is fixed at $\rm RV_{H\alpha}=0\,km\,s^{-1}$. We Doppler shift the spectrum of a B star (dotted lines) by $\pm 53\,\rm km\,s^{-1}$, representing the luminous star's orbital motion. This causes the total spectrum (emission plus absorption; solid lines) to shift in the opposite direction. We then fit for the mean velocity of the H$\alpha$ emission at each epoch using the wings of the total emission profile, without accounting for contamination from the absorption line. This causes the emission line to appear to shift by $\pm 4.4\,\rm km\,s^{-1}$, exactly in anti-phase with the luminous star. This shift is not real; the emission line's velocity is fixed at 0 by construction.}
    \label{fig:schematic}
\end{figure*}

In this Letter, we reconsider the evidence for the mass of the unseen companion in LB-1. We argue that the companion is probably a stellar-mass BH, but that there is no evidence for an unusually high mass.

\section{Basic issue}
\label{sec:issue}

The observed spectrum of LB-1 contains absorption lines, presumably formed in the photosphere of a B star, and emission lines (most notably, H$\alpha$), with some other origin. B stars have strong H$\alpha$ absorption lines with broad wings, so the observed flux at the H$\alpha$ line is a sum of emission and absorption (i.e., a reduction in the stellar continuum). The stellar absorption lines are known to be RV-variable. If one wishes to determine the velocity of the H$\alpha$ {\it emission} alone, it is necessary to first account for the wavelength-dependent reduction in total observed flux due to the absorption line. This is illustrated schematically in Figure~\ref{fig:schematic}, where we show the effects of an RV-variable absorption line on the apparent velocity of an emission line whose true velocity is constant. 

The emission line is modeled with a Voigt profile centered on 0 with Gaussian $\sigma=90\,\rm km\,s^{-1}$, Lorentzian $\gamma = 50\,\rm km\,s^{-1}$, and amplitude 2.8. As we discuss in Section~\ref{sec:data}, we find this functional form to provide a good fit to the Keck/HIRES H$\alpha$ emission profile of LB-1, particularly its wings. We model the spectrum of the B star with a TLUSTY model \citep{Hubeny_1995} with $T_{\rm eff}= 18,000$\,K, $\log g=3.5$, and $Z=Z_{\odot}$, consistent with the fit to the spectrum of the B star obtained by \citet{Liu_2019}\footnote{Several works \citep[e.g.][]{AbdulMasih_2019, Simon_Diaz_2019, Irrgang_2019} have reported a cooler temperature, $T_{\rm eff}\approx 13,000$\,K. None of our results change significantly if we adopt this value.}. We broaden the spectrum to $v\sin i=10\,{\rm km\,s^{-1}}$, as found for the observed B star, using the rotational broadening profile from \citet{Gray_1992}. 
We shift the stellar spectrum by $\pm 53\,\rm km\,s^{-1}$, representing changes in the velocity of the B star throughout its orbit, and then add the emission line (always centered on $\rm RV_{\rm H\alpha}=0\,km\,s^{-1}$) to the shifted stellar spectrum at each epoch. We degrade the total spectrum to the resolution of the Keck/HIRES data analyzed in the next section ($R\approx 60,000$), assuming a Gaussian line spread function. The details of our treatment of the spectrum do not sensitively affect our results, because the H$\alpha$ absorption line is broad and basically featureless in the region used to measure its velocity.

\begin{figure*}
    \includegraphics[width=\textwidth]{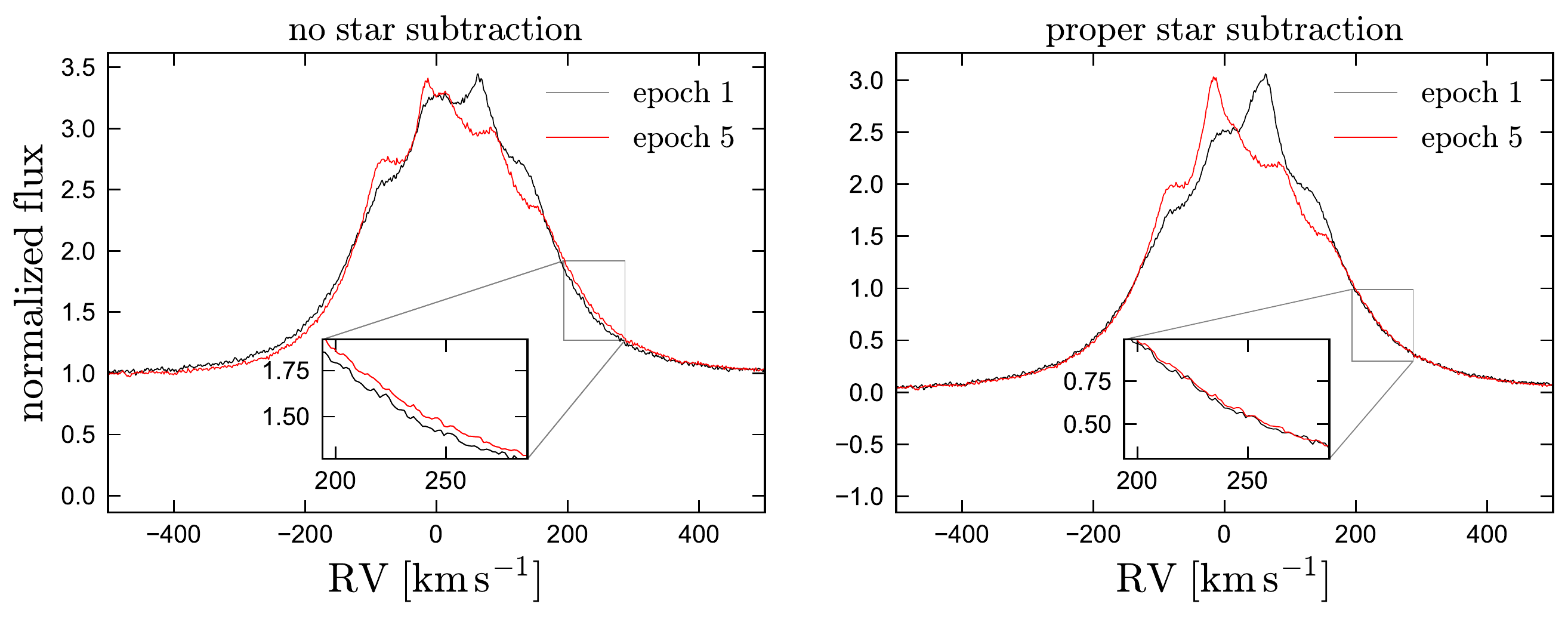}
    \caption{Left: comparison of Keck/HIRES spectra of LB-1 obtained at two different epochs, centered on the H$\alpha$ line. The velocity of the B star varies by $71\,\rm km\,s^{-1}$ between these epochs. The wings of the emission line profiles also suggest a velocity offset between them, and fitting yields an offset of $10\,\rm km\,s^{-1}$. Right: the same two spectra after the template spectrum of a B star at the appropriate velocities has been subtracted. There is now no evidence of a velocity offset between the two emission lines: the apparent offset in the left panel was due to the RV variability of the stellar absorption line.}
    \label{fig:shifted_spectra}
\end{figure*}

Figure~\ref{fig:schematic} shows that shifts of the absorption line cause the total emission profile to shift in anti-phase with the star. When we fit for the line's mean velocity (using the profile wings, as described in the next section), we infer an erroneous shift of $\pm 4.4 \rm\,km\,s^{-1}$, {\it even though the line is stationary by construction}. When the stellar spectrum is subtracted from the total emission, we of course recover the true $\rm RV_{\rm H\alpha}=0\,km\,s^{-1}$ at all epochs. One might hope that masking the line center during fitting would protect against contamination from absorption. It does not, because although the absorption line core is narrow and excluded from the region where RVs are fit, the Balmer lines in the B star have broad wings, such that the magnitude of emission and absorption in LB-1 are comparable at $\left |\rm RV \right| \gtrsim 200\,\rm km\,s^{-1}$.

\section{Contamination from absorption in LB-1}
\label{sec:data}

We now investigate the effect of contamination from H$\alpha$ absorption on the spectrum of LB-1. We analyze the same 7 epochs of Keck/HIRES data analyzed by \citet[][their Extended Data Table 1]{Liu_2019}. The data were obtained and reduced using the standard California Planet Search setup \citep{Howard_2010} and are available through the Keck Archive. We focus on the Keck/HIRES data because it has the highest spectral resolution, allowing the most precise measurement of the H$\alpha$ line's velocity, but we expect measurements based on spectra from other instruments (e.g. LAMOST and GTC) to be affected in the same way.

We barycenter-corrected the spectra and normalized them by dividing by the mean flux at $6580 < \lambda/{\rm AA} < 6600$; this simple procedure works well because the spectra are featureless in this region and we only study a narrow wavelength range centered on H$\alpha$.   

Figure~\ref{fig:shifted_spectra} compares the H$\alpha$ line profiles of LB-1 in the two epochs between which the velocity of the B star varies most. In the left panel, we simply overplot the normalized spectra from the two epochs. Here, there appears to be a visible shift between the two line profiles, and fitting for the emission line velocities (see Section~\ref{sec:fitting_rv}) yields a velocity shift of $10\,\rm km\,s^{-1}$ between them. However, the right panel of Figure~\ref{fig:shifted_spectra} shows that once the TLUSTY spectrum of the B star is subtracted, the line profile wings in the two epochs become almost perfectly aligned, with no evidence for a velocity shift between them. We find very similar results when comparing all pairs of single-epoch spectra.

\subsection{Fitting RVs}
\label{sec:fitting_rv}

\begin{figure*}
    \centering
    \includegraphics[width=\textwidth]{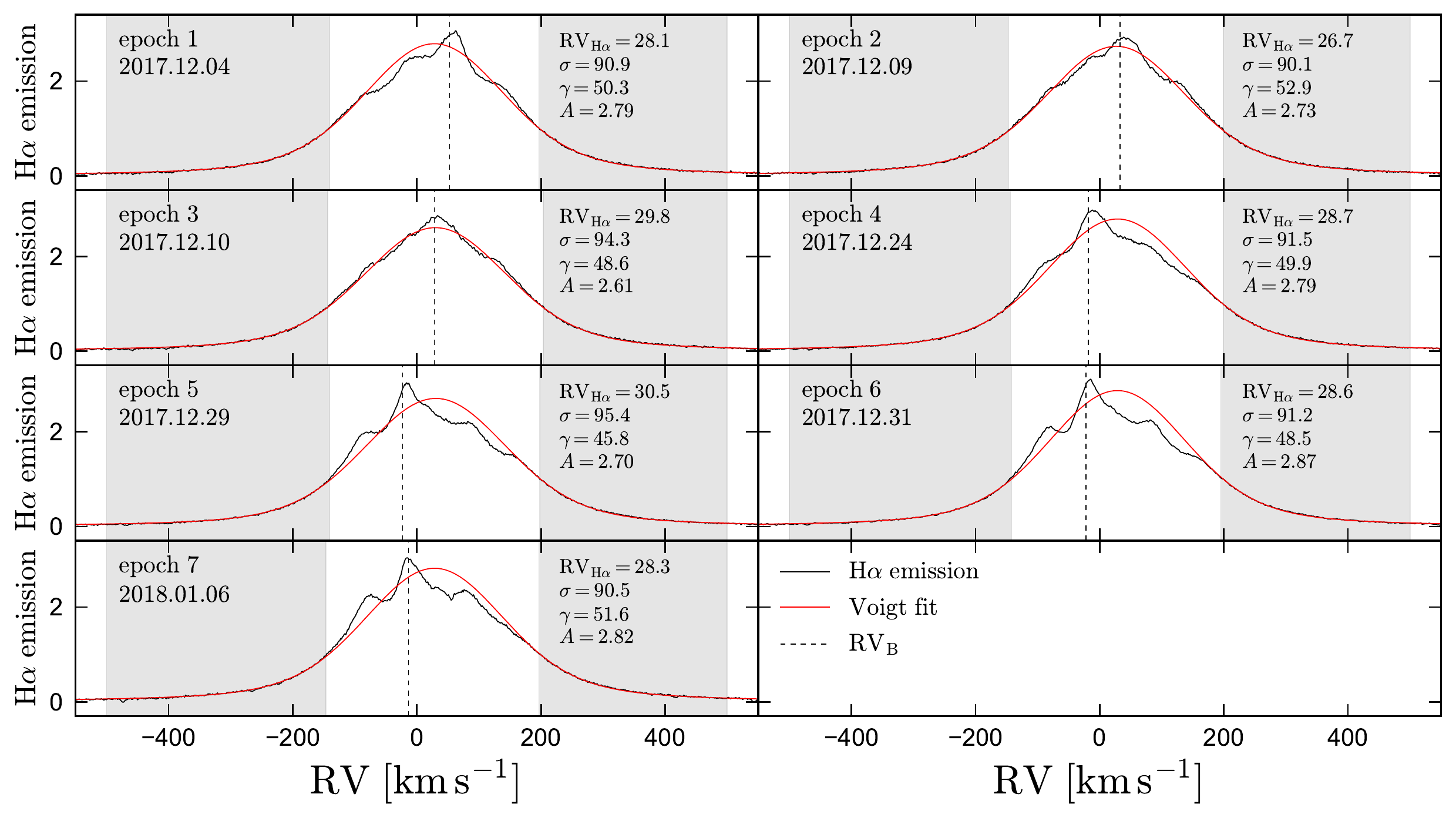}
    \caption{Voigt profile fits to all 7 epochs of Keck/HIRES data, after subtraction of the stellar spectrum. Gray shaded regions show the data used in fitting; the line centers are masked. Although the shape of the line center varies between epochs, the wings are relatively stable and can always be reasonably well-fit by a Voigt profile. Best-fit parameters are listed for each epoch. Dashed vertical lines show the velocity of the B star as measured from absorption lines. A narrow component of the emission appears to track the B star.}
    \label{fig:fits}
\end{figure*}

We now turn to measuring the radial velocity of the H$\alpha$ emission line at all epochs. As shown in Figure~\ref{fig:shifted_spectra}, the emission near line center (within $\pm 150\,\rm km\,s^{-1}$ of the mean velocity) varies substantially between epochs, but the shape of the emission line wings is relatively stable. Following \citet{Liu_2019}, we measure the line velocity using the region of the spectrum with $\rm |RV| < 500\,\rm km\,s^{-1}$, while masking out the line center, which is defined as the region where the emission line flux exceeds 1/3 of its maximum value (shaded regions in Figure~\ref{fig:fits}). We measure the velocity of the line at each epoch by fitting a Voigt profile, which we find can always provide a good fit to the line profile within these windows.

Each fit has 4 free parameters: a central velocity $\rm RV_{H\alpha}$, an amplitude $A$ (which we define as the profile's value at line center), a Gaussian dispersion $\sigma$, and a Lorentzian HWHM $\gamma$. Figure~\ref{fig:fits} shows the results of these fits. We also carry out similar fits {\it without} first subtracting the spectrum of the B star. The formal fitting uncertainty in $\rm RV_{H\alpha}$ is typically $(0.1-0.2)\,\rm km\,s^{-1}$. We inflate this uncertainty by adding to it a $1\,\rm km\,s^{-1}$ systematic error in quadrature to account for uncertainty in the wavelength solution when it is not registered to lines of known wavelength \citep[e.g.][]{Griest_2010}. Our fitting procedure is not identical to the one used by \citet{Liu_2019}, but we find it to be robust and show below that it yields consistent results. 

\begin{figure}
    \includegraphics[width=\columnwidth]{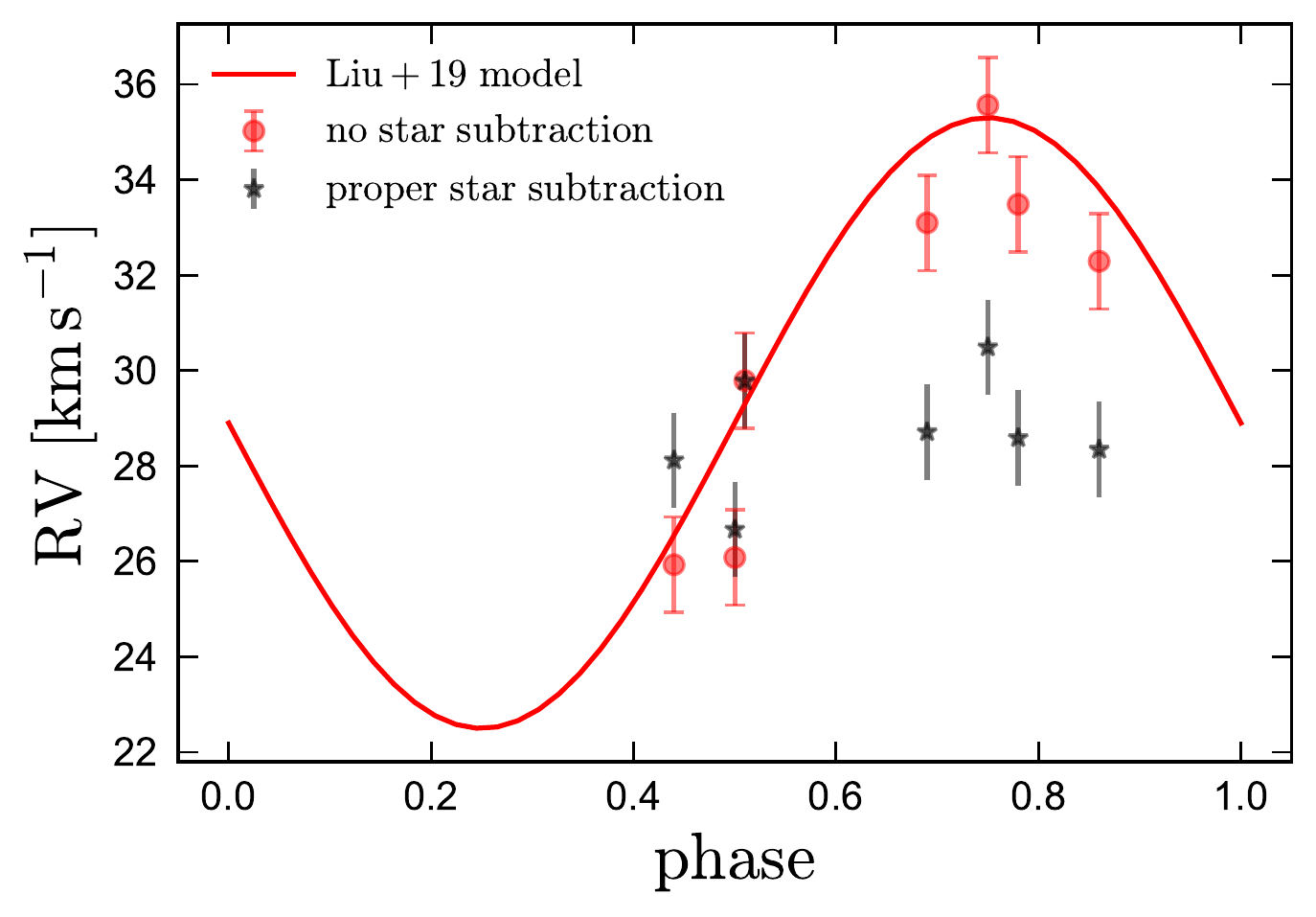}
    \caption{Red points show RV measurement of LB-1's H$\alpha$ emission line obtained from Keck/HIRES spectra without accounting for the stellar absorption line (e.g. left panel of Figure~\ref{fig:rv_curve}). These show clear evidence of RV variability and are consistent with the orbital solution for the H$\alpha$ line derived in \citet{Liu_2019}. Black points show RV measurements once contamination from stellar absorption lines is accounted for (e.g. right panel of  Figure~\ref{fig:rv_curve}). Here there is no evidence of RV variability, and the model with $K_{\rm H\alpha}=6.4\,\rm km\,s^{-1}$ is manifestly ruled out.  } 
    \label{fig:rv_curve}
\end{figure}

In Figure~\ref{fig:rv_curve}, we plot the results of this RV fitting as a function of the binary's orbital phase. Red points show RV measurements when contamination from the B star's absorption lines is not corrected for. As expected, these are in agreement with the model from \citet{Liu_2019}, which has $K_{\rm H\alpha}=6.4\,\rm km\,s^{-1}$, and they are inconsistent with a constant velocity for the H$\alpha$ line. Black points show RV measurements after the B star template spectrum has been subtracted. These show no evidence for variation with orbital phase and appear consistent with a constant velocity for the H$\alpha$ line wings. 

The data are strongly inconsistent with $K_{\rm H\alpha}=6.4\,\rm km\,s^{-1}$ and in fact rule out any $K_{\rm H\alpha}> 1.3\,\rm km\,s^{-1}$ at the 2-sigma level. The best-fit sinusoidal velocity amplitude is formally $K_{\rm H\alpha} = 0.25^{+0.7}_{-0.3}\,\rm km\,s^{-1}$, but a constant velocity provides essentially as good a fit as the best-fit sinusoidal model ($\Delta \chi^2 = 0.1$). Given the sinusoidal model's higher complexity, there is no evidence that the H$\alpha$ line is RV-variable. 

The dashed vertical lines in Figure~\ref{fig:fits} show the velocity of the B star at each epoch. Although the wings of the emission line are stationary, there is a narrow component of the emission ($\rm FWHM \approx 50\,km\,s^{-1}$) that shifts between epochs and appears to trace the B star. Moreover, the emission flux in the inner wings (outside the narrow peak) is correlated with the velocity of the star: in epochs 4-8, when the star is blueshifted relative to the barycenter, there is excess emission in the left wing relative to the Voigt fit, and a deficit of emission in the right wing. The situation is reversed in epoch 1, when the star is redshifted relative to the barycenter.

\section{Discussion}
\label{sec:discussion}
We have shown that the H$\alpha$ line in LB-1, the binary reported to contain a $70\,M_{\odot}$ BH, does not actually display evidence of periodic RV variability, at least not at the $K_{\rm H\alpha } > 1.3\,\rm km\,s^{-1}$ level. This undermines the interpretation of the previously reported $K_{\rm H\alpha}$ in terms of the system mass ratio, and thus the reported 70 $M_{\odot}$ mass. 

\subsection{Nature of the unseen companion}

One conceivable interpretation is that the companion is a BH with even higher mass than reported: assuming $M_{B}=8.2$, our upper limit on $K_{\rm H \alpha}$ would imply $M_2 \gtrsim 330 M_{\odot}$. We regard this scenario as exceedingly unlikely both because of the astrophysical challenge of producing such a system and because it would imply an improbably low inclination ($i \lesssim 9\,\rm deg$, representing $\sim$1\% of randomly oriented orbits). We thus conclude that the H$\alpha$ line (at least its wings) is not moving with either component of the binary. 

Although there is little evidence that LB-1 contains an unusually massive BH, the unseen companion may still be a stellar-mass BH. If the mass of the B star is indeed $M_B \approx 8 M_{\odot}$ as found by \citet{Liu_2019}, this would imply $6\lesssim M_{2}/M_{\odot}\lesssim20 $ for inclinations $30<i/{\rm deg}<90$ (representing 87\% of randomly oriented orbits). Normal-star companions in this mass range are expected to contribute to the spectrum at a detectable level, so the lack of absorption lines from a luminous secondary would make a ``normal'' stellar-mass BH the most likely companion. 

Two additional observations point toward the companion being a stellar remnant. First, the orbit is nearly circular. This is rare for ordinary main-sequence binaries with $P \gtrsim 10\,{\rm days}$, which have a wide range of eccentricities \citep{Moe_2017, Elbadry_2018}, but it is expected for binaries containing evolved stars, because even binaries with relatively long periods ($P \lesssim 200\,\rm days$) become tidally circularized when one component ascends the giant branch \citep{Verbunt_1995, PriceWhelan_2018}. If LB-1 was tidally circularized when the companion was a giant, then the fact that the orbit is still circular may set stringent limits on any mass loss or kicks that occurred during the death of the companion. 

Second, the rotation velocity of the luminous component ($v\sin i \approx 10\,\rm km\,s^{-1}$) is exceptionally low for a B star. This could be explained if the star was tidally synchronized when the companion was a giant, such that the rotation period is equal to the orbital period. Synchronization is expected to occur on a timescale similar to or shorter than circularization. The expected rotation velocity in this case is $v_{{\rm rot}}=2\pi R_{B}/P$, which ranges from 4.5 to 7\,$\rm km\,s^{-1}$ for $7<R_{B}/R_{\odot}<11$, similar to the observed rotation velocity. If the B star is tidally synchronized, then a precise measurement of $v\sin i$ and the stellar radius could constrain the orbital inclination.

As discussed by \citet{Liu_2019}, their spectroscopically-inferred $M_B\approx 8 M_{\odot}$ is inconsistent with the {\it Gaia} DR2 parallax, which implies a lower mass, $M_B\approx 5 M_{\odot}$. The revised atmospheric parameters estimated by \citet{Simon_Diaz_2019} and \citet{AbdulMasih_2019} also support such a scenario, which would yield a lower limit of $M_2 \gtrsim 5\,M_{\odot}$, still making a BH the most likely companion.

\citet{Eldridge_2019} and \citet{Irrgang_2019} proposed that the luminous star could be a $\sim 1 M_{\odot}$ pre-subdwarf in a short-lived ($\sim 10^4 \rm \,yr$) evolutionary phase. In this scenario, the companion could also have a significantly lower mass: for $M_{B} < 1.2 M_{\odot}$, the minimum mass of the companion is less than $2.5 M_{\odot}$, such that it could be a neutron star. The primary challenge to this scenario is its short lifetime: the probability is low that the LAMOST RV monitoring campaign, which observed $\sim 3000$ targets, would happen to observe such a rare object. 

\subsection{Source of the H$\alpha$ emission}
The shape of the H$\alpha$ profile near line center varies with orbital phase, with a narrow component that tracks the velocity of the B star (Figure~\ref{fig:fits}). This may originate at the star itself, perhaps from accretion or from interaction of the stellar wind with the circumbinary environment. It is possible that a component of the emission tracks the BH, but this is not easily distinguishable with the presently available spectra.

As for the wings, the fact that their mean velocity does not vary with phase suggests that this emission must not originate from either component of the binary, but rather from material in the barycentric rest frame. Because the present-day orbit is smaller than the radius of plausible progenitors to the companion when it was a giant, the system likely went through a period of common envelope evolution. Fallback material from the companion's envelope could form a circumbinary disk in this scenario \citep[][]{Muno_2006, Kashi_2011}. If the luminous star is a recently-stripped pre-subdwarf \citep[e.g.][]{Irrgang_2019}, fallback material from its envelope could also contribute. Another possibility is that the low-velocity component of winds from the B star are captured by the binary \citep[e.g.][]{Taam_2001}.

For a $5 M_{\odot}$ BH and a $5 M_{\odot}$ star, $a\approx 0.75$\,AU, and the circular velocity at the inner rim of a circumbinary disk ($r \approx 2a$) would be $v_c \approx 75\,\rm km\,s^{-1}$. For an optically thin Keplerian disk, one would thus expect a double-peaked emission line with peaks separated by $\sim 150\,\rm km\,s^{-1}$. The H$\alpha$ line is somewhat wider than this, with $\rm FWHM \approx 260\,\rm km\,s^{-1}$ and shallow wings that extend at least $\pm 500 \rm km\,s^{-1}$ from line center. The wings likely do not trace disk kinematics, but are instead broadened by a combination of non-coherent scattering (i.e., repeated absorption and re-emission of line photons in an optically thick medium; see \citealt{Hummel_1992}) and electron scattering (in the outermost wings; see \citealt{Poeckert_1979}). The effects of these processes in broadening H$\alpha$ are well-studied in the disks around classical Be stars \citep[e.g.][]{Hanuschik_1996}. Spiral density waves, interactions with the B star's winds, and fluctuations of the disk's rim temperature due to the B star's orbital motion can all complicate the emission line profile. In the future, a more dense sampling of spectra in orbital phase may enable reconstruction of the emission region and more detailed modeling of the circumbinary environment. 

There are several similarities between LB-1 and the recently-identified binary AS 386 \citep{Khokhlov_2018}, which contains a $7\,M_{\odot}$ B-star primary and a detached $\gtrsim 7 M_{\odot}$ unseen companion suspected to be a BH. That system also has a strong H$\alpha$ emission line whose shape changes between epochs. Although the mean velocity of the system's H$\alpha$ emission does vary with phase, its velocity amplitude is lower than that of the star, and Doppler tomography suggested that it originates from a shock where the wind from the B star interacts with a circumbinary disk. The orbit of the B star in AS 386 is also circular, despite a long period of 131 days. Like LB-1, and in contrast to most classical Be stars, AS 386 is slowly rotating. Also in contrast to classical Be stars \citep{Dachs_1990}, both LB-1 and AS 386 have stronger Balmer decrements than predicted for low-density nebular gas, case-B recombination, and a temperature near $10^4\,\rm K$. The spectral energy distributions of AS 386 and LB-1 both reveal infrared excess; \citet{Khokhlov_2018} find that in AS 386, this stems primarily from dust emission in a circumbinary disk. 

AS 386 exhibits periodic variability in the infrared. \citet{Khokhlov_2018} attribute this to dust temperature variations in the Earth-facing inner rim of the circumbinary disk as the distance from the B star to the rim changes. We do not find evidence of such variability in the WISE light curves of LB-1, although the light curves are sparse. This could indicate that the inclination is lower than in AS 386, or that the IR excess (which is weaker than that in AS 386) is emitted at larger radii. We also note that unlike LB-1, the spectrum of AS 386 contains several other emission lines, including forbidden lines emitted in the circumbinary disk. Further theoretical modeling of LB-1 will be required to explain these differences.

Although it likely does not contain a $70\,M_{\odot}$ BH, LB-1 is an intriguing system which adds to the short but growing list of Milky Way stellar mass BH candidates that are not X-ray bright \citep[e.g.][]{Casares_2014, Khokhlov_2018, Giesers_2018, Thompson_2019}. If the unseen companion is confirmed to be a BH, LB-1's circular orbit will provide tantalizing evidence that some massive stars end their lives without an explosion \citep[e.g.][]{Adams_2017}. We are optimistic that further study of the system will shine additional light on its evolutionary history.

\section*{Acknowledgements}
We thank the anonymous referee for a constructive report, and JJ Eldridge, Andrew Howard, Howard Isaacson, Andreas Irrgang, Chris Kochanek, Jifeng Liu, Ben Margalit, Todd Thompson, Dan Weisz, Yanqin Wu, and Wei Zhu for helpful discussions.
K.E acknowledges support from an NSF graduate research fellowship.



\bibliographystyle{mnras}




\bsp	
\label{lastpage}
\end{document}